\documentclass[reprint,amsmath,amssymb,aps,prb]{revtex4-2}
\usepackage[dvipsnames]{xcolor}
\definecolor{red}{RGB}{219, 1, 1}
\usepackage{graphicx}
\usepackage{dcolumn}
\usepackage{bm}

\begin{document}

\title{Band Structure and Charge Ordering of Dirac Semimetal EuAl$_{4}$ at Low Temperatures} 

\author{Andrew Eaton$^{1,2}$, Brinda Kuthanazhi$^{1,2}$, Paul. C. Canfield$^{1,2}$, Benjamin Schrunk$^{1,2}$, Na Hyun Jo$^{3}$, Yevhen Kushnirenko$^{1,2}$, Evan O’Leary$^{1,2}$, Lin-Lin Wang$^{1,2}$ and Adam Kaminski$^{1,2}$}

 \email{adamkam@ameslab.gov}

\affiliation{%
 $^{1}$Ames Laboratory, U.S. Department of Energy, Ames, Iowa 50011, USA \\ 
 $^{2}$Department of Physics and Astronomy, Iowa State University, Ames, Iowa 50011, USA \\
 $^{3}$Department of Physics, University of Michigan, Ann Arbor, Michigan 48109, USA}%

\date{\today}

\begin{abstract}
EuAl$_{4}$ is proposed to host topological Hall state. This material also undergoes four consecutive antiferromagnetic (AFM) transitions upon cooling below T$_{N1}$= 15.4 K in the presence of charge density wave (CDW) order that sets in below T$_{CDW}$= 140 K. We use angle-resolved photoemission spectroscopy (ARPES) and density-functional-theory (DFT) calculations to study how magnetic ordering affects the electronic properties in EuAl$_{4}$. We found  changes in the band structure upon each of the four consecutive AFM transitions including band splitting, renormalizations and appearance of  new bands forming additional Fermi sheets. In addition we also found significant enhancement of the quasiparticles lifetime due to suppression of spin flip scattering, similar to what was previously reported for ferromagnetic EuCd$_2$As$_2$. Surprisingly, we observe that most significant changes in electronic properties occurs not at T$_{N1}$, but instead at the AFM3 to AFM4 transition, which coincides with largest drop in resistivity.
\end{abstract}
\maketitle
\section{INTRODUCTION} 
Rare-earth compounds containing Eu attract attention because their divalent and trivalent electronic states and highly dense 4f electron bands lead to interesting physical properties of the materials that contain them\cite{Jo2020, Onuki2013,Onuki2014}. Eu in EuAl$_4$ is divalent and this material has been the focus of several studies for exhibiting unique electronic properties and several magnetic orderings.\cite{Onuki2013, Onuki2014, Onuki2015, Kobata2016, Onuki2019, Kaneko2021, Shang2021, Takagi2022, Meier2022, Ramakrishnan2022, Zhu_2022, Gen2023}. Centrosymmetric, tetragonal EuAl$_4$ crystallizes in BaAl$_4$ structure\cite{Onuki2019} that belongs to the \emph{I}4/\emph{mmm} No. 139 space group shown in Fig. 1(a). Recent studies report EuAl$_4$ to be the first example of a centrosymmetric binary compound where stacking corrugated layers of non-magnetic Al and magnetic square layers of Eu enables this system to host skyrmion spin textures\cite{Shang2021, Takagi2022, Meier2022}. Skyrmions and other topological spin textures have attracted much interest because their tunable small size has the potential to significantly reduce the size of next generation electronic devices\cite{Back2020, GOBEL20211, Tokura2022}.

\begin{figure*}
\includegraphics[scale=0.6715]{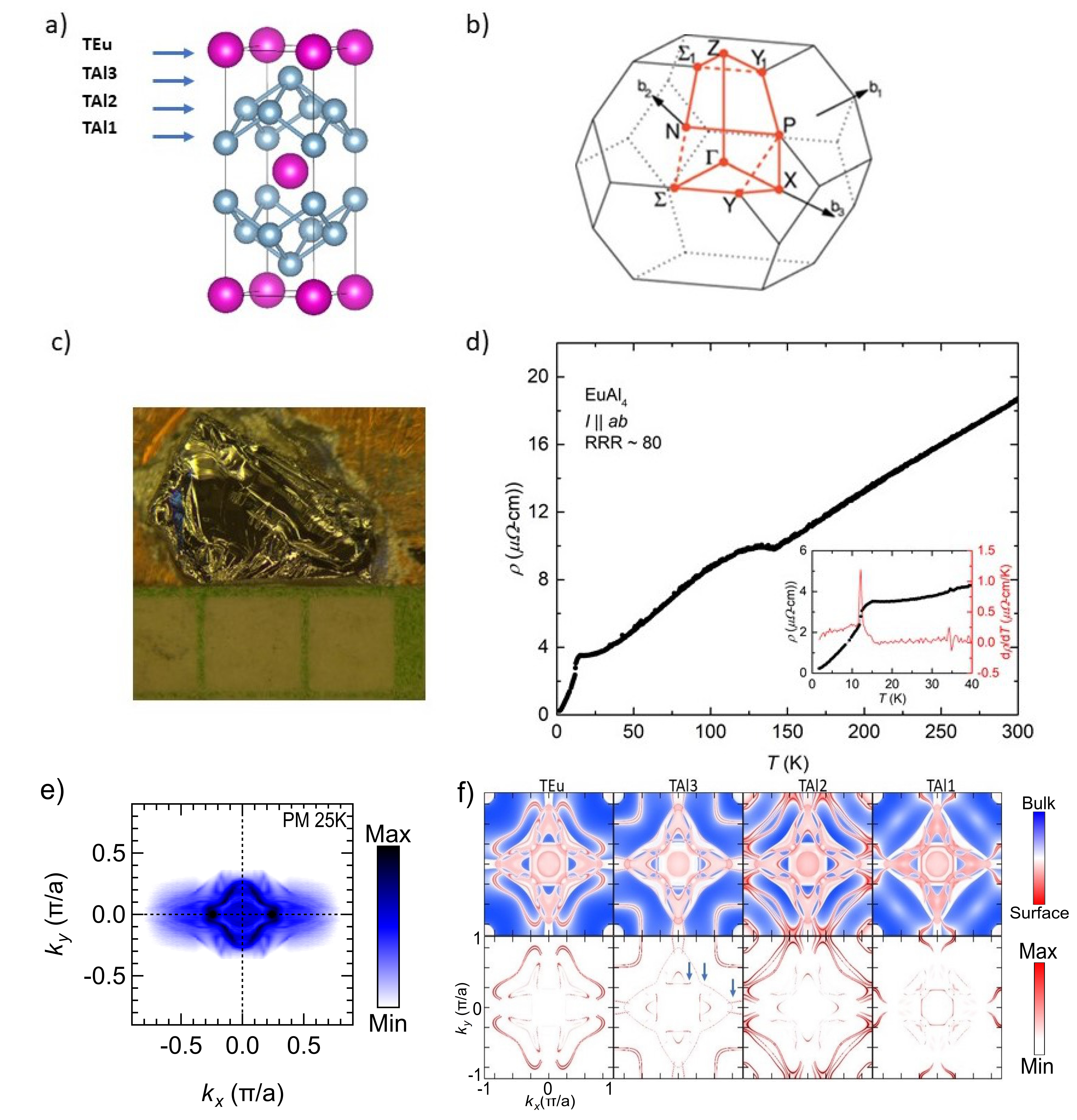}
\caption{{a) Conventional unit cell of EuAl$_4$ with termination planes marked. b) First Brillouin zone of EuAl$_4$. c) Single crystal of EuAl$_4$ after cleaving and measurement with 1mm x 1mm grid paper for scale. d) Resistivity  measurements as a function of temperature. In-lay in corner shows resistivity in expanded temperature scale below 40K in black and  its first derivative  in red. e) Symmetrized PM state Fermi surface at 25K. f) DFT calculations of the first Brillouin zone for the four termination planes of EuAl4. The top row shows combined surface and bulk bands, while only surface states are shown in the bottom row.}}
\end{figure*}

Previous studies reported the presence of charge density waves (CDW) below T$_{CDW}$, where electron-phonon interactions spontaneously cause a charge density modulation and a periodic lattice distortion\cite{Onuki2019, Ramakrishnan2022, Gen2023}. EuAl$_{4}$ presents a rare case where CDW and antiferromagnetic (AFM) ordering co-exist. In the paramagnetic state (PM) at ambient pressure below T$_{CDW}$= 140 K, a CDW order develops on the Al atoms\cite{Onuki2014, Onuki2014, Onuki2015, Kaneko2021, Shang2021, Ramakrishnan2022}. Plausible evidence for an orthorhombic CDW transition has been presented in temperature-dependent electrical resistivity measurements and diffraction experiments\cite{Nakamura2013, Onuki2013, Onuki2014, Onuki2015, Kobata2016, Kaneko2021,Ramakrishnan2022}. At lower temperatures, AFM order develops with CDW present and four AFM transitions are reported to occur at T$_{N1}$= 15.4 K, T$_{N2}$= 13.2 K, T$_{N3}$= 12.2 K, and T$_{N4}$=  10.0 K\cite{Onuki2013, Onuki2015, Onuki2019, Kaneko2021}. EuAl$_4$ is unique in that it preserves the CDW order after crossing the first Neel temperature, which persists upon cooling through successive AFM orderings\cite{Meier2022}. The coexistence of these symmetry-breaking phenomena attracts interest, as both are dependent on the Fermi surface through Fermi surface nesting and Ruderman-Kittel-Kasuya-Yosida (RKKY) interactions between the localized magnetic moments of the 4f electrons\cite{Onuki2015, Kobata2016}. Neutron diffraction analysis of these AFM orderings reveals differences in the propagation vectors\cite{Kaneko2021}. The AFM transitions at T$_{N1}$ and T$_{N2}$ are of $\lambda$-type second order variety with magnetic diffraction peaks indexed at (q, q, 0) and (q, -q, 0)\cite{Onuki2015, Kaneko2021}. The T$_{N3}$ and T$_{N4}$ AFM transitions are first order, as evident by the temperature dependent steep drop in the specific heat with (q, 0, 0) and (-q, 0, 0) propagation vectors\cite{Onuki2015,Kaneko2021}. Lattice distortions are reported below T$_{N3}$, as a tetragonal to orthorhombic structural phase transition occurs.\cite{Onuki2019, Gen2023}. Real spin-space maps for the different AFM orderings were reported demonstrating EuAl$_{4}$'s AFM ordered states contain spin textures, including the formation of skyrmion spin lattices at AFM2 and AFM3 orderings\cite{Gen2023}. Previous angle-resolved photoemission spectroscopy (ARPES) studies of EuAl$_4$ utilized a synchrotron source and were confined to a temperature range between 20 K and 200 K\cite{Kobata2016}. In this study, we probe the electronic structure of EuAl$_4$ using laboratory based Ti:sapphire laser ARPES at temperatures down to 7 K. From this, we mapped the evolution of the electronic structure upon all four successive AFM transitions. The crystal's termination plane was determined by comparing the data taken in the PM state with density-functional-theory (DFT) calculations.

\begin{figure*}
\centering
\includegraphics[scale=0.55]{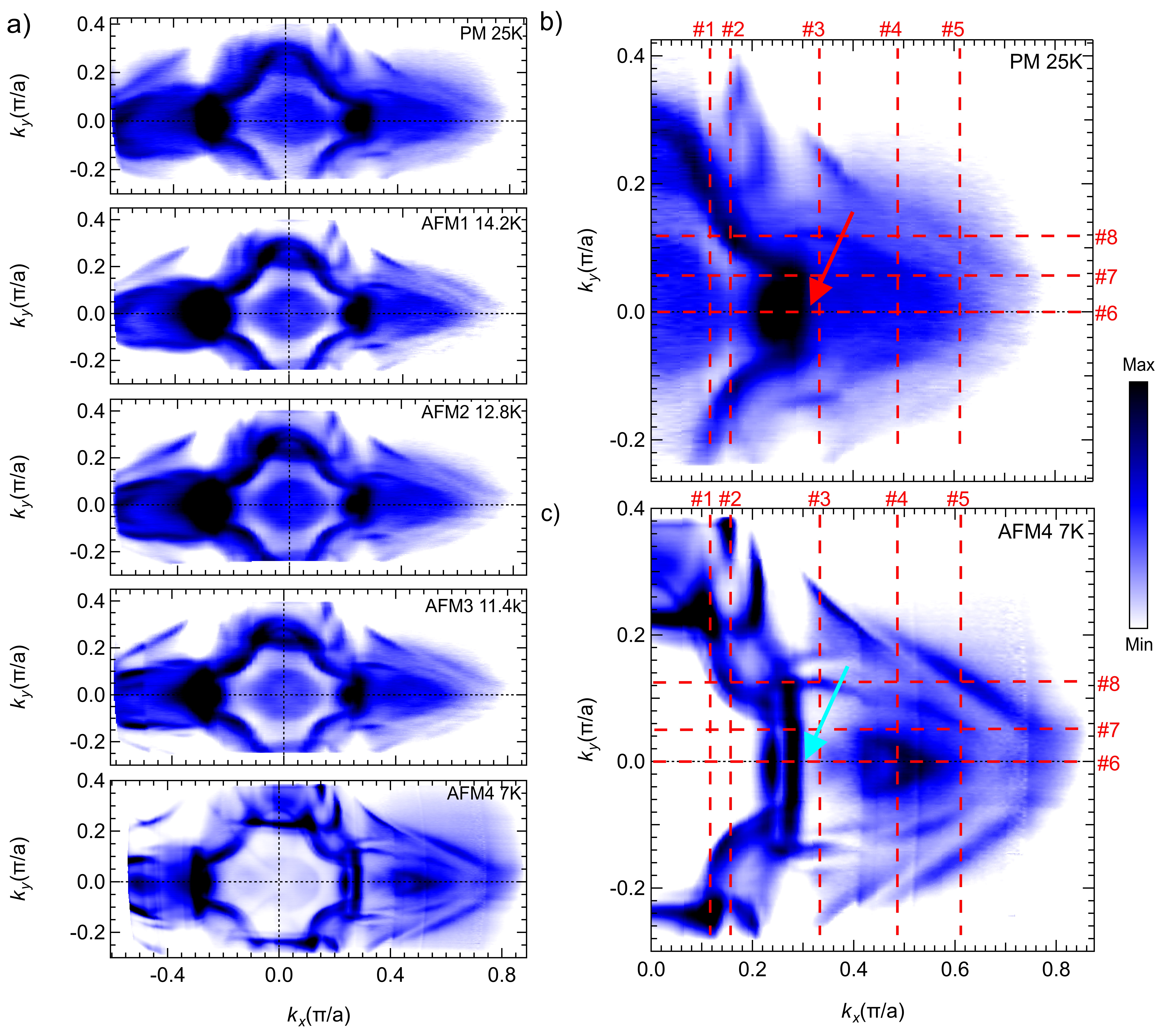}
\caption{a) Fermi surfaces maps taken in between each magnetic transition, labeled in the corner of each map. b) Fermi surface at 25 K zoomed into region of analysis with cuts shown in Fig. 3 and 4 marked. c) Fermi surface at 7 K zoomed into region of analysis with cuts shown in Fig. 3 and 4 marked.}
\label{fig:Marked Cuts}
\end{figure*}

\section{EXPERIMENTAL DETAILS}
Single crystals of EuAl$_{4}$ were grown out of Al flux. The elements with an initial stoichiometry of Eu$_{8}$Al$_{92}$ were put into a Canfield fritted alumina crucible \cite{Canfield2016} and sealed in fused silica tube under partial pressure of argon. Such prepared ampoule was heated to 900$^{\circ}$C over 5 hours and held there for 2 hours. This was followed by a slow cooling to 660$^{\circ}$C over 100 hours and decanting of the excess flux using a centrifuge \cite{Canfield_2020}. Temperature dependent resistivity data shown in Fig. 1(d) agree well with reported behavior, showing a dip associated with CDW ordering at 140 K and drop-off at low temperature. The inlet of Fig. 1(d) shows the low temperature regime of the resistivity in black and the derivative of the resistivity with respect to temperature d$\rho$/dT in red. 

\begin{figure*}
\centering
\includegraphics[scale=0.5]{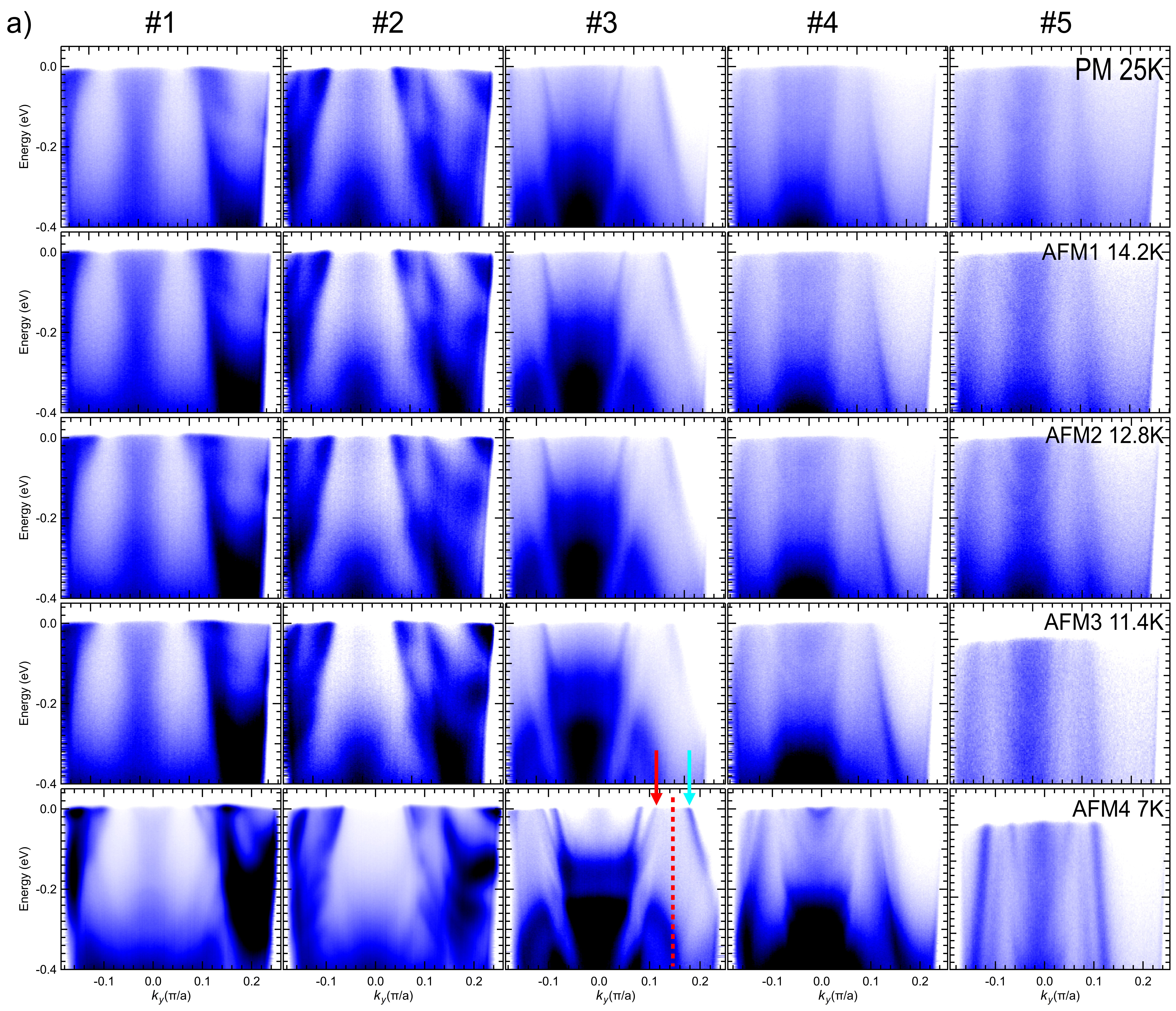}
\caption{a) ARPES vertical cuts $\#$1-5 corresponding to marked numbers measured between each magnetic state with temperature decreasing from top to bottom. Rows are in the same state, marked in the corner of the right panel. Cut $\#3$ in the AFM4 state is marked to show the potential magnetic zone boundary. The left arrow shows potential band back folding about the zone boundary of the band marked by right arrow.}
\end{figure*}

Single crystals were cut to size, mounted on a copper puck, and cleaving pins were attached to the top (001) surface. Fig. 1(c) shows the sample of EuAl$_4$ after measurements on a 1mm grid paper for scale. ARPES measurements were performed using a laboratory-based vacuum ultraviolet (VUV) laser ARPES spectrometer that consists of a picosecond Ti:sapphire oscillator with fourth harmonic generator, low temperature cooling stage with a pulse tube cryostat, and a Scienta DA30 hemispherical electron analyzer\cite{jiang2014tunable, Schrunk2019, yunwu}. All data were collected with a photon energy of 6.7 eV. Angular resolution was set at approximately ~0.1$^{\circ}$ and 1$^{\circ}$ along and perpendicular to the direction of the analyzer slit, respectively, and the energy resolution was set at 2 meV. The size of the vertically polarized photon beam spot on the sample was  $\sim$15 $\mu$m. The sample was cleaved \textit{in situ} at UHV pressure below 2 $\times$10$^{-11}$ Torr. Data were collected in EuAl$_4$'s PM state above the Neel temperature and while in each successive AFM state below T$_{N1}$. Data for Fermi surfaces maps were taken at 7 K, 11.5 K, 12.9 K, 14.3 K, and 25 K, respectively, with temperatures stabilizing prior to each measurement.

\begin{figure*}
\centering
\includegraphics[scale=0.45]{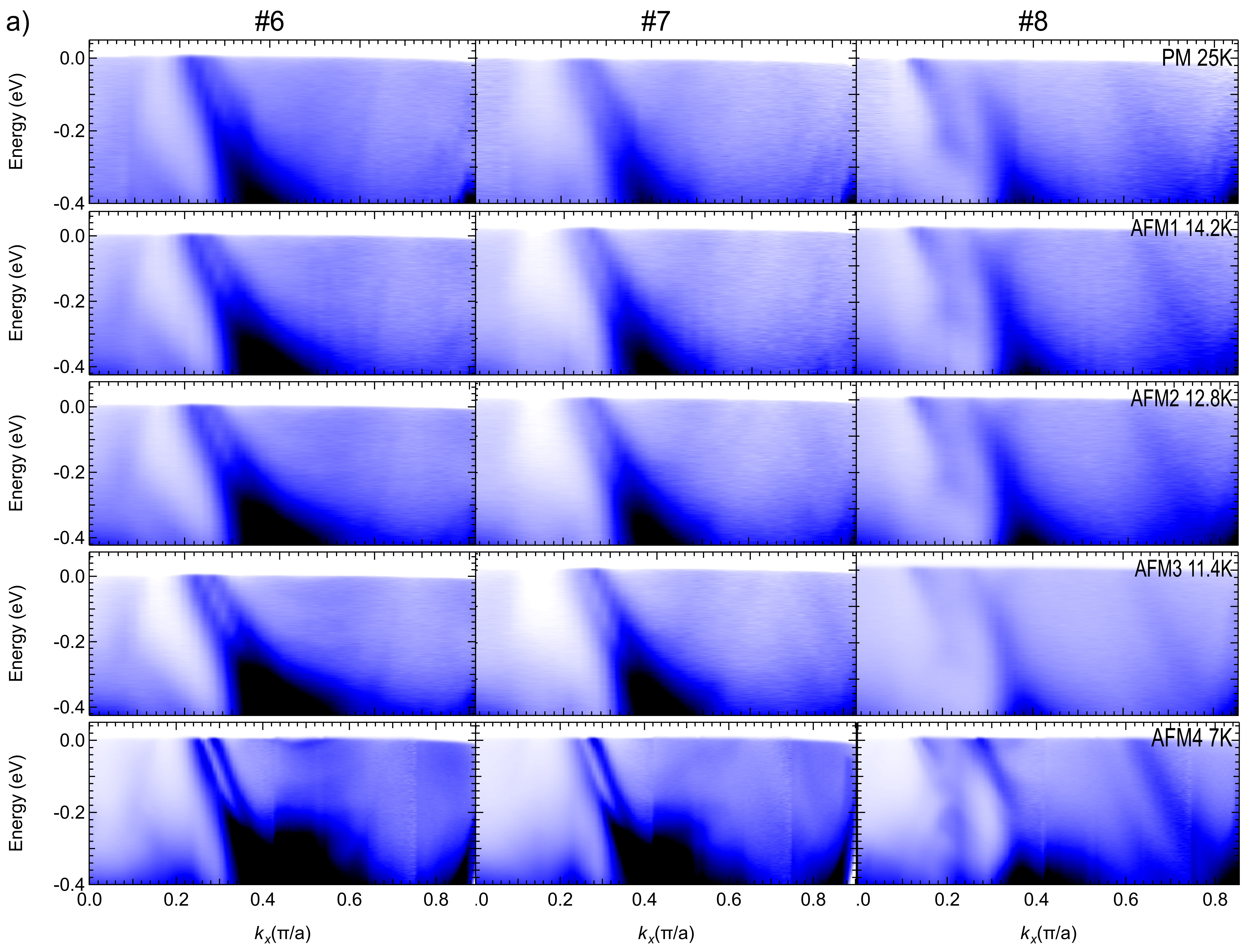}
\caption{a) ARPES horizontal cuts $\#$6-8 made along $\Gamma$-$\Sigma$ direction taken at decreasing temperature from top to bottom measured between each magnetic state. Rows are in the same state, marked in the corner of the right panel.}
\end{figure*}

DFT calculations were performed using Vienna Ab initio Simulation Package (VASP)\cite{Kresse1996,KRESSE199615} assuming non-magnetic state of the sample (Eu's 4f as core electrons). Data in Fig. 1(e) are symmetrized around horizontal and vertical axis to make the 25 K PM state Fermi surface shown. Fig. 1(f) shows the intensity map calculations of the bulk plus surface states in the top row and only the surface states in the bottom row. 

\begin{figure*}
\centering
\includegraphics[scale=0.2715]{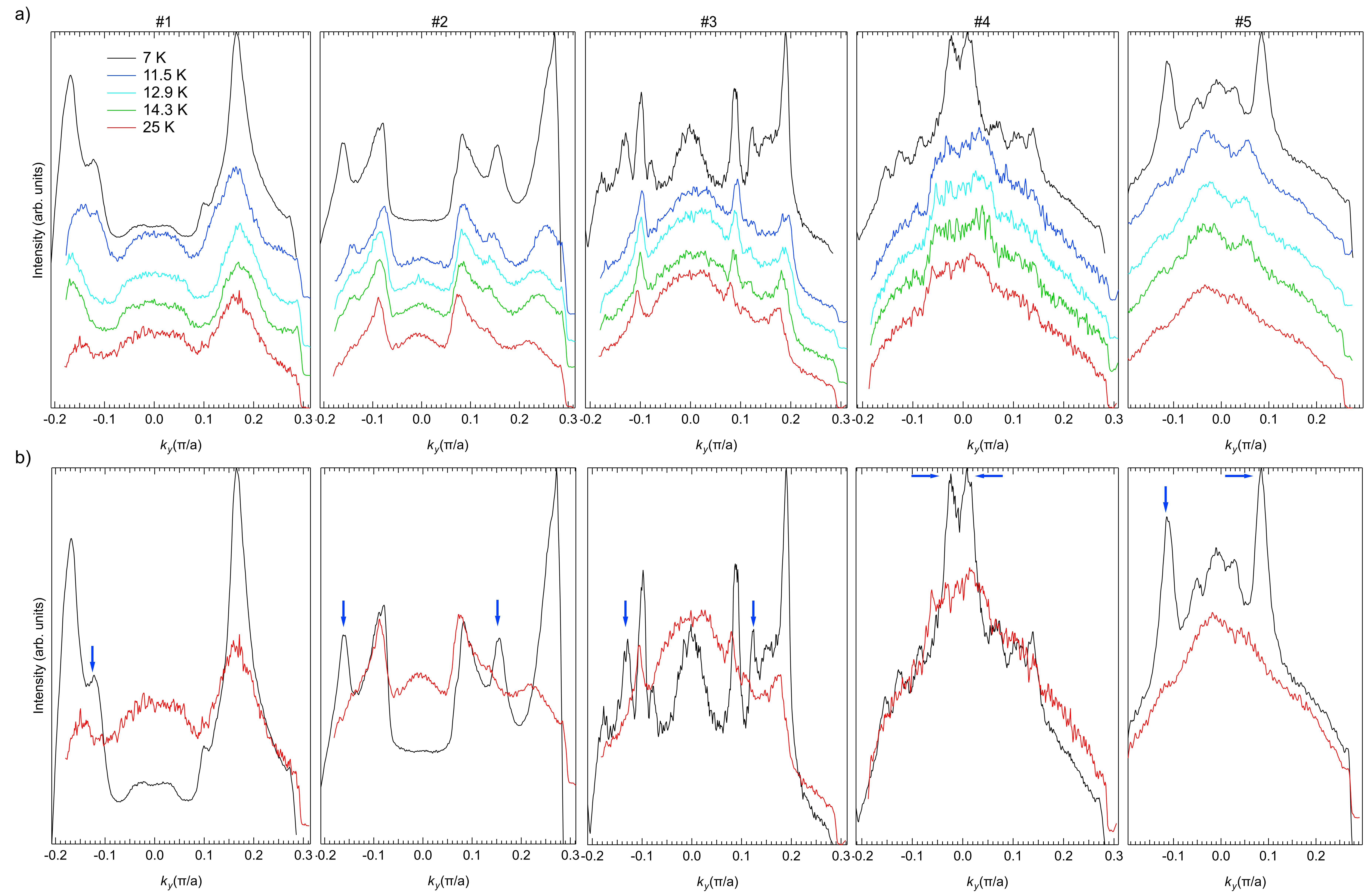}
\caption{a) Momentum distribution curve (MDC) corresponding to the FS cuts in Fig. 3, for each temperature. Cuts are labeled above each MDC. The bottom (red) MDC is the PM state and temperature decreases with each MDC to the lowest order AFM state at the top (black). b) Comparison of MDCs from the PM state (red) and lowest AFM ordered state (black). Blue arrows indicate bands that experience significance enhancement in the AFM4 state.}
\end{figure*}

\section{RESULTS AND DISCUSSION}
We performed DFT calculation to better understand the data measured by ARPES and determine the most likely cleaving plane. In all experiments we always observed the same bands structure and Fermi surface, which indicates the presence of a strongly preferred cleaving plane, where the bonds are weakest and where sample fracture occurs. By comparing the symmetrized  Fermi surface measured in the PM state in Fig. 1(e) with DFT calculations carried out for the four most likely cleaving planes in Fig. 1(f), we analyzed which calculated surface termination is the best agreement with our data. We note the high intensity and sharp bands around $\Gamma$ that appear straight with edges that curve away from $\Gamma$ around \textit{k$_{x}$} and \textit{k$_{y}$}=$\pm$~0.25 $\frac{\pi}{a}$ in Fig. 1(e). This band likely corresponds to the surface state located in the same position in the bottom row of Fig. 1(f) for termination TAl3, which is marked with blue arrows. Additional evidence for the TAl3 termination is found near the edge of the first Brillouin zone (BZ), where two straight bands begin at \textit{k$_{x,y}$}= $\pm$~0.75 $\frac{\pi}{a}$ and appear to continue to the edge of the first BZ. We note that in the DFT calculations, only the TAl3 termination has surface bands that extend out of the first BZ in this way without additional bands crossing or terminating in the region above \textit{k$_{x,y}$}=$\pm$~0.75 $\frac{\pi}{a}$. The sharp outer band that curves toward these parallel bands corresponds to the "flower" shape band marked with the middle blue arrow in the surface state of TAl3. Based on these observations, we conclude that the TAl3 termination site is the most likely cleaving plane in our experiments. Since we did not observe any variation of the data for different cleaves or locations on the sample surface, we concluded that this material cleaves with TAl3 termination.

\begin{figure*}
\includegraphics[scale=.4]{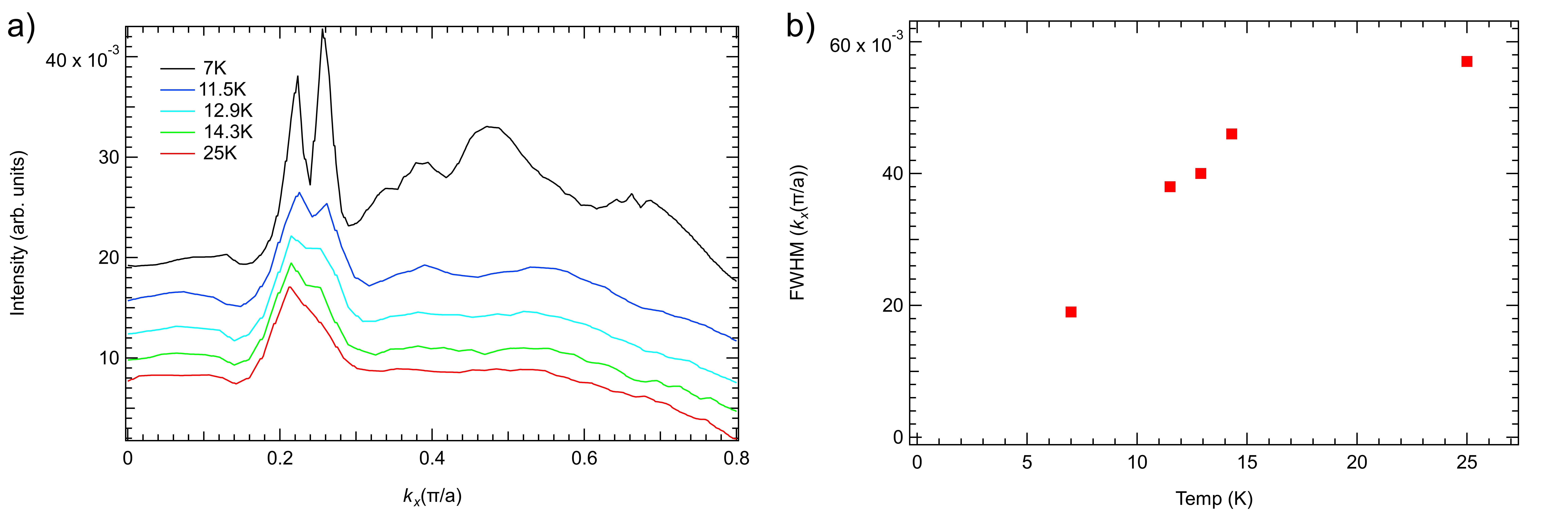}
\caption{a) MDCs from cut $\#$6 taken at the Fermi level. b) FWHM of the MDCs, showing a decrease in the FWHM at lower temperatures and the largest drop occurring in the AFM4 state.}
\end{figure*}

We show the temperature evolution of portions of the Fermi surface around $\Gamma$ in Fig. 2. The Fermi surface in the PM and in the antiferromagnetically ordered AFM1-4 states are shown in Fig. 2(a). Enlarged portion of the Fermi surface in the PM state and AFM4 are shown in Fig. 2(b) and (c), respectively for more detailed comparison. We notice there are several electron pockets and bands in the Brillouin zone  appearing in the AFM4 phase that were not present in the PM phase. Most notably, a new horizontal segment of the Fermi surface appear at \textit{k$_{y}$}=$\pm$~0.24 $\frac{\pi}{a}$ on Fig. 2(c). An arrowhead shaped pocket also appears at \textit{k$_{x}$}=0.5 $\frac{\pi}{a}$ about the horizontal symmetry direction. The nearly parabolic Fermi surface sheet at \textit{k$_{x}$}=$\pm$0.25 $\frac{\pi}{a}$ splits into shorter and longer, almost flat segments. In addition, significant sharpening of other bands occur. Most of the above changes occur only upon transition to the AFM4 phase, as seen in Fig. 2(a), that is differences between PM and AFM1-3 are smaller by comparison than difference between data in AFM3 and AFM4 phases at lowest temperatures. This is consistent with transport measurements, where the largest changes in resistivity occur slightly above 10 K, i.e., near AFM3 to AFM4 transition as seen in the inset of Fig. 1(d).

To better understand the evolution of the Fermi surface with temperature, we analyzed several horizontal and vertical cuts in the BZ in more detail. The ARPES intensity data representing the band dispersion in the PM and AFM1-4 states along a series of vertical cuts $\#$1-5 are shown in Fig. 3. A very broad electron band at momentum \textit{k$_{y}$}= 0 $\frac{\pi}{a}$ located at the center of cut $\#$1, is present in PM and AFM1-4 states and does not seem to be affected by any of the transitions. Other hole-like broad bands are present at momentum \textit{k$_{y}$}= $\pm$0.15 $\frac{\pi}{a}$ that become slightly sharper upon cooling below the AFM2 transition and appear to split into two separate bands. Upon cooling below the AFM4 transition, cut $\#$1 shows further sharpening of these broad bands with a new band present at momentum \textit{k$_{y}$}= 0.2 $\frac{\pi}{a}$ that does not appear in the data at higher temperatures. Cut $\#$2 shown in Fig. 3 has several broad hole bands present near the Fermi energy in the PM state. We do not observe any significant changed in the band dispersion, nor the width of the bands, apart from small changes in relative intensity. It seems that quasiparticles in this cut are not significantly affected by any of the AFM transitions. Cut $\#$3 shown in Fig. 3 has several broad bands present in the PM state that sharpen when temperature is decreased below T$_N$. We note the emergence of a sharp band at momentum \textit{k$_{y}$}= $\pm$0.13 $\frac{\pi}{a}$ in AFM4 (marked by red arrow) that was not present at higher temperatures. The PM state in cut $\#$4 shown in Fig. 3 has broad bands present that do not change significantly upon decreasing temperature to AFM1-3 states. When cooling below the AFM4 transition, a sharp electron pocket at \textit{k$_{y}$}= 0 $\frac{\pi}{a}$ that was unresolvable in the data at higher temperatures and perhaps even absent, becomes sharper. This results in higher peak intensity, indicating an increase of the quasiparticle lifetime. Cut $\#$5 shown in Fig. 3 has several broad bands that form the 4-fold "flower" structure of the PM state Fermi surface. These bands remain mostly unchanged upon cooling below the AFM1-3 state transitions. Decreasing temperature below the AFM4 state transition causes a significant reduction in the peak width, an increase of intensity, and splitting of this band. 

The momentum distribution curves (MDCs) at the Fermi energy for vertical cuts are shown in Fig. 5(a).  The bottom MDC was measured in the PM state, while top MDC is from AFM4 state. Each MDC was plotted with an offset for  clarity. We compare directly the MDCs in the AFM4 state to ones in the PM state for each cut in Fig. 5(b). The MDCs in the PM state in cut $\#$1 show the presence of the band at the center as well as another band at momentum \textit{k}= $\pm$0.2 $\frac{\pi}{a}$. Upon decreasing the temperature, a small hump can be seen in the MDCs that appears below the AFM2 transition. As temperature is decreased further below the AFM3 transition, a new peak begins to emerge at the location of the hump and the intensity of this band increases when the sample is further cooled below the AFM4 transition, which is marked in left panel of Fig. 5(b) by a blue arrow. The data along cut $\#$2 in Fig. 5(a) shows the MDCs for the PM state with broad bands present near the Fermi energy, including the hole pocket as noted in previous paragraph. As temperature is decreased below the AFM1-3 transitions, the intensity corresponding to the hole pocket at the center decreases and this feature vanishes once the sample is cooled below the AFM4 transition. In contrast, the peaks corresponding to bands marked with blue arrows in cut $\#$2 of Fig. 5(b) emerge out of background upon cooling. Cut $\#$3 in Fig. 5(a) shows the broad bands in the PM state experience intensity enhancement as the sample is cooled into AFM1-3 states. Upon further cooling below the AFM4 transition, new bands emerge that were not present in AFM1-3, as seen from the presence of peaks at momentum \textit{k}= -0.09 $\frac{\pi}{a}$ and \textit{k}= $\pm$0.12 $\frac{\pi}{a}$ marked in Fig. 5(b).  Cut $\#$4 in Fig. 5(a) shows the broad bands that compose the outer 4-fold "flower" pattern of the PM state Fermi surface are present upon cooling below the AFM1-3 transitions. In the AFM4 state, large double peak feature emerges in the data along cut $\#$4 centered at the symmetry direction axis. The MDCs along cut $\#$5 in the PM and AFM1-3 states are rather broad and mostly featureless. Upon cooling below the AFM4 state transition, five resolvable peaks appear in the data as seen in right panels of Fig. 5(a) and (b). 

Interesting changes in the electronic properties occur along horizontal Fermi surface cuts, the ARPES intensity for which is plotted in Fig. 4. In the AFM4 state of cut $\#$6, an electron pocket is present near E$_{f}$ centered at \textit{k$_{x}$}=0.5 $\frac{\pi}{a}$ which vanishes upon increasing temperature and is no longer visible in the AFM3 state. The very broad band present at \textit{k$_{x}$}=0.25 $\frac{\pi}{a}$, forms a hole-like pocket that splits into two upon cooling below T$_{N}$, while both branches remain relatively broad. They become much sharper once the AFM3 to AFM4 transition occurs. We document the changes in the width of the MDC peaks along cut $\#$6 in Fig. 6. In Fig. 6(a), we plot the MDC in PM as well as each of the AFM states. A broad peak visible in the PM state appears to slightly split when sample is cooled to AFM3 state. Then quite suddenly, the width of the two split peaks decreases upon further cooling to AFM4 state and two sharp and clearly visible peaks are present. We extracted the full width half maximum (FWHM) of the peaks and plot their temperature dependence in Fig. 6(b). There is initial decrease of FWHM below T$_N$, but the largest drop occurs in the AFM4 state, which is consistent with data discussed above and drop in resistivity in Fig. 1(d).

It is clear from the MDCs that dramatic changes occur upon the series of AFM transitions, with the largest changes occurring for the AFM3 to AFM4 transition. These include the appearance of the new peaks in the AFM4 phase, which signifies the emergence of new bands, disappearance of others, and decrease in the width of the peaks due to enhanced lifetime of the quasiparticles. The first change is due to the formation of new magnetic zone boundaries and back folding of the existing bands from other parts of the Brillouin zone. Examples of this can be seen in cut $\#$3 in Fig. 3, where a new band appears (marked by a red arrow) that is a reflection of the existing band (marked by blue arrow) about the potential magnetic zone boundary. The disappearance of  some of the bands is quite surprising, but can be potentially explained by the emergence of hybridization gaps due to band back folding. Other effects reported here are not easily explained by common understanding of evolution of band structure upon AFM transitions. A good example of this is the change of rather broad and semi-round portions of the Fermi surface sheet that is present in the PM, AFM1, 2, and 3 states in Fig. 2(b) (marked by red arrow), into a long and almost straight arc in the AFM4 state in Fig. 2(c) (marked by blue arrow). Unfortunately, understanding of these effects via DFT calculations requires knowledge of the exact magnetic structure of each phase that is not currently available. The enhancement of quasiparticle lifetime occurs due to suppression of spin flip scattering, when long range magnetic order is established. In the PM state, there is very strong scattering of quasiparticles due to large and disordered local magnetic moments of Eu ions. Once long range magnetic order is established, the contribution of the magnetic moments  to the Hamiltonian becomes periodic and magnetic scattering is suppressed. The additional periodicity leads to formation of new magnetic zone boundaries and band back folding as described above. The enhancement of quasiparticle lifetime, leads to significant decrease of the resistivity, as seen in transport data shown in Fig. 1(d). It is interesting that the largest changes in resistivity occur for the AFM3 to AFM4 transition, which coincides with the largest changes in the band dispersion and quasiparticle lifetime reported here. We foresee that similar effects should occur in other Eu and Gd based antiferromagnets due to their large, local magnetic moment. We also note that similar enhancement of the quasiparticle lifetime was previously reported in ferromagnetic EuCd$_{2}$As$_{2}$ with both, however different types of magnetically ordered states sharing very similar electronic response.

\section{CONCLUSION}
We presented ARPES data demonstrating how the electronic properties of EuAl$_{4}$ change upon the emergence of four subsequent AFM orders. We established that the experimental cleaving plane is consistent with the TAl3 termination by comparing DFT calculated Fermi surfaces for four possible terminations with the measured Fermi surface in the PM state. We report significant changes in band dispersion and quasi-particle lifetime upon the series of AFM transitions. These changes are a combination of sharpening of bands sensitive to the crystals magnetic structure, emergence of new bands upon entering lower temperature ordered states, and bands disappearing from the Fermi surface after a transition. The remaining bands experience significant changes of their quasi-particle lifetime, which is most likely linked to suppression of spin flip scattering when magnetic moments of Eu develop long range order. While we have established changes that occur to the Fermi surface structure for EuAl$_{4}$'s PM and AFM ordered states upon low temperature evolution, more study is needed to fully understand the affect CDW and magnetic ordering interplay has on EuAl$_{4}$'s electronic properties.

\section{ACKNOWLEDGMENTS}
This work was supported by the Center for the Advancement of Topological Semimetals, an Energy Frontier Research Center funded by the U.S. Department of Energy Office of Science, Office of Basic Energy Sciences through the Ames Laboratory under its Contract No. DE-AC02-07CH11358.

\hbadness11000
\providecommand{\noopsort}[1]{}\providecommand{\singleletter}[1]{#1}%

\end{document}